# QoS Web Service Security Dynamic Intruder Detection System for HTTP SSL services

## M.Swami Das[1], A.Govardhan[2], D.Vijaya lakshmi [3]

Assoc. Professor, CSE, MREC[1], Professor, SIT,JNTU Hyderabad[2], Professor, Dept. of CSE, MGIT Hyderabad, India[3]

**Abstract:** Web services are expected to play significant role for message communications over internet applications. Most of the future work is web security. Online shopping and web services are increasing at rapid rate. In this paper we presented the fundamental concepts related to Network security, web security threats. QoS web service security intrusion detection is important concern in network communications and firewalls security; we discussed various issues and challenges related to web security. The fundamental concepts network security XML firewall, XML networks. We proposed a novel Dynamic Intruder Detection System (DIDA) is safe guard against SSL secured transactions over message communications in intermediate routers that enable services to sender and receiver use Secured Session Layer protocol messages. This can be into three stages 1) Sensor 2) Analyzer and 3)User Interface..

1.**Introduction**
Intrusion detection system is a device or software application that monitors malicious attacks or network traffic if any policy violations [1]. Web services applications communicate and coordinate message passing between client and server. A web service provides functionality and services to the web users. The users to communicate in network channels, the hacker or Intruder tries to operate various attacks such as, DDOS attack, side channel attack, authentication attack, man in the middle attack, cloud computing attacks and steal sensitive information. Hacker execute arbitrary or malicious code in the system due to vulnerability, weak security and no Intruder Detection and Monitoring system.[2]. In Intrusion detection system is a device or software application that monitors malicious attacks or network traffic if any policy violations [1]. A web service provides functionality and services to the web users. The users to communicate in network channels, the hacker or Intruder tries to operate various attacks such as, DDOS attack, side channel attack, authentication attack, man in the middle attack, cloud computing attacks and steal sensitive information. Hacker execute arbitrary or malicious code in the system due to vulnerability, weak security and no Intruder Detection and Monitoring system.[2]. In recently ISRO website homepage hacked by hackers, other examples related to Government and other web sites discussed[3,4]. It is essential to provide Intruder Detection and monitoring system for Government Institutions, Diplomatic offices, Energy, oil and gas companies, Research Institutions, private equity firms, and activist. Frequently to monitor and control the valid and authorized data operations over the network.

Web security has three important concepts confidentiality, integrity and availability. Confidentiality means Information not available to unauthorized users. Integrity defined by the property that data has not been modified by unauthorized users, and availability means web services are accessible to authorized users with access restrictions.[5] Intrusion: attempting to attack into or misuse the system from outside network or legitimate users of the network, intrusion can be a physical, system or remote intrusion. Automatic Intrusion detection system sensor, Analyzer and user interface. Intrusion Detection systems can be classified as i)Anomaly detection ii) signature based misuse iii) host based iv) network based v) stack based

The rest of this paper is organized as follows Section 2: Related work, Section: 3 Issues and challenges, Section 4: Web security and Network security, Section: 5 Discussions and Interpretations and Section 6. Conclusion.

## 2. Related Work

Zhiwen Bai etl, Proposed DTAD, a dynamic taint analysis detector aiming to protect malicious attacks and vulnerabilities. Attacker process is detected and precision intrusion, signature of collection of virtual systems and comparing network data and log files used to identify the attacks.[6]

Jiang Du etl, studied man in the middle attacker use ARP deception for both sides communication. Man in the middle will generate own public, private and self digital certificate, and this is interactive process validated by Service provider.[7]

Taro Ishitaki,etl proposed intrusion detection system using Neural network, Fuzzy logic, Probabilistic reasoning, Genetic algorithms capable for finding pattern behavior to detect normal and attack conditions[8].

The SOAP messages to ensure integrity and authentication during the data transmission. Web services require partial signing of SOAP request which is achieved using XML signature by WSDL documents and operations as suggested by Padmanabhuni and Adarkar etl[9,10]. Web service security is critical task for message invocations by web servers. SOAP uses XML encryption, XML digital signature, SSL/TLS methods. XML message security is achieved by service oriented security functionality.

Web service standards SOAP level security authentication, authorization management. Web security is defined as attach signature and encryption header to SOAP messages. It describes security tokens. Web security policy is defined as set of specifications that describe rules, constraints and other business policies on intermediaries and end points. (Example. Encryption algorithms).Web security trust describes a frame work to design a model that enables web services to securely inter-operate request, issues, and exchange security operations.

### 2.1 XML Firewall

Web services environment, malicious attacks and DoS attacks are new challenges. Firewall allows to the Service providers residing in a network to be invoked from outside the network, and keeping a high security[9]. HTTP protocol is not suitable for creating public key infrastructure. The prototype is used by application behind a firewall.[11]

### 2.2 XML networks

web service management vendors develop network based solution for web service applications to provide better QoS web services with security to various networks endpoints service consumers and service providers.[11].

Fang Qi .etl, .proposed Automatic Detecting Security Indicator (ADSI) for preventing Web spoofing on a confidential computer which is a harmless environment. It creates a random indicator to identify and detect bogus pages with URL screening data.[13]

Jaing Du,.etl analyzed as a case study secured socket layer man in the middle attack based on SSL certification interaction. Attacker place computer gives a vital role two communication processes. [14]

Lin-Shung Huang., etc introduced a new method for detecting SSL man -in- the middle attacks against website users, over of SSL connections at the top web sites by checking certificates as number of CA certificates. Trace any malware in SSL connections for identify and provide better protection. [15]

## 3 Issues and challenges

The following are the list of issues /challenges in Web security/network security. Digital certificates are designed to establish credentials of the people use Router configurations with weak vulnerabilities and security policies described in Table.1. Web security developers provide secured operations and safety steps necessary to identify trusted systems. [16]

Table.1. Router or firewall configurations with weak or vulnerabilities

| Web services Solutions or threats | Problem in domain or Safety precautions |
|---|---|
| Web service has arbitrary disclosure policy | Provide strong policies to web services |
| Passwords stored in browser | Do not save passwords in browser history |
| Institutions, organizations malicious code attacks, virus | Web security , Frequently monitor network operations. Use SSL security |

Malware, Denial of service attacks to modems / routers against other systems by unknown users by stealing personal information and credentials to access certain web sites.

Hackers used stolen laptops/equipment to hack web data where there is vulnerability like private wireless network or wireless network is unsecured with no password is immediately accessible to hackers. Hacker used wireless antenna and software nearby buildings and capture/ steal information like passwords, email-messages, and any data transmitted over the network when a network is not secured. [5].Hacker will use some tools described in table.2. Brute-force attack is the password cracking method, trying all the solutions seeking one fits[11].Stealing the login password controlling the devices by malicious scripts and malicious DNS servers attacked on DSL modems.

### 3.1 Man-in-the-Middle (MIM) attacks

This attack where the attacker secretly relays and possibly alters the communication between two parties who believe they are openly communicating with each other. Attacker intercept all relevant messages by passing between victims and adding extra information. Attackers trying to access the services using fake address, fake certifications. Examples of MIM attacks One provides free Wi-Fi service with malicious software.

#### 3.1.1 ARP Cache Poisoning

Sender and receiver over message communication, PC sends IP packets broad cast to all systems in subnet. ARP(address resolution protocol is not secured protocol).

#### 3.1.2 DNS Spoofing

DNS cache poisoning is a computer hacking attack, where by data is communicated into a Domain Name System (DNS) resolver's cache, causing the name server to return an incorrect IP address, diverting traffic to the attacker's computer (or any other computer).Attackers creating a fake web site by redirecting data to shadow servers.

#### 3.1.3 Session Hijacking:
Client to server when session established, the hacker capture cookies information and diverting the session communications to un-trusted systems

#### 3.1.4 Session hijacking attack

Communication over TCP connections. Session normally consists of string of variables used in URL stealing and predicting valid session token to gain unauthorized access to the web server [17]

### Table.2.Tools and software's used to steal the data

| Web services Solutions or threats | Problem in domain or Safety precautions |
|---|---|
| Suspicious downloads or plugins | Use firewall in secure network |
| Terminals with chip card vulnerabilities | Alert any where service by authentication and secret key. |

### 4.Web Security Network security

Web Service Security: Three types of digital certificates are domain validated certificate, organizational validation certificate and Extended validation certification. Domain validated Certificate: trusted domain name of owner. Organizational Validation Certificate: validation of organization by DNS names. Extended validation certification: Certificate Agent must meet minimum validation criteria. Organizations, application vendors, Browser makers issue extended validation certificate.[4] Web services standards worked at w3C, OASIS, IETF and other bodies to enable faster inventions of web services and security. A web service provides a flexible set of mechanism to design a range of security protocols. It is essential to design non-vulnerable protocols for web services security. Web services specifications goals to provide multiple security token formats, multiple trust domains, multiple signature formats, multiple encryption methodologies, and end to end message content security.[12]

### 4.1 Intruder Identification and Detection System

#### *4.1.1 Various Attacks*

Unauthorized system used to attack on router or servers using various attacks (DDOS attack, side channel attack, Man in the middle attack , Authentication attack and cloud computing attacks) methods practiced due to various reasons like, not secured web site, malicious code, denying encrypt , weak secret keys, vulnerabilities in content security, and policy constraints. In Figure.1. shows the intruder attacks on router.

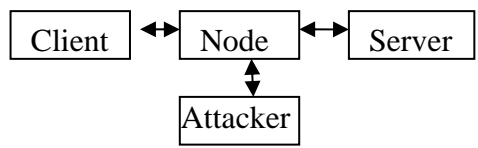

Figure.1. Intruder attacks on services

*4.1.1.1 Denial of Service attacks*
DDOS attack were launched from distributed attacking hosts. This is launched two phases. First an attacker builds a network which is distributed and consists of thousand of compromised computers are called(Zombies, attacking hosts). The attacker hosts flood of tremendous volume of traffic towards victims either under command or automatically [32].

*4.1.1.2 Attack to change DNS settings*
Attackers directly targeting DNS server two ways Cybersquatting aim is to steal the Victims identity and or divert traffic from victims website. Name jacking or theft : by appropriate the domain name (updating the holders field or taking control) by technical means to divert the traffic such as modifying the name of hosting the site.[18]

*4.1.1.3 Authentication attack*
This type of attack targets and attempt to take advantage of following Brute force : allow attacker to guess persons username, other credentials by using Automated trail and error Insufficient Authentication: Allows an attached to access a web site sensible information without having to properly authenticate in web site. Sending phishing mail to user to steal sensitive information[19]

**4.2 Intruder Detection System**
Intruder Detection System has two type namely Network Intrusion Detection System and Host based Intrusion Detection System

*4.2.1 Network based Intrusion Detection System*
It deals with traffic accounting and network flow information. This system is implementing in Routers and switches Input and Output HTTP / TCP data, and testing various functions like port scanning , Reassembling, decoding, detecting virus, protocol violations.

*4.2.2 Host based Intrusion Detection System*
It deals with Analyzing logging facility for almost all failed or success services. The system is implementing in Routers or Firewall to access authorized client. It calculates the cryptographic checks of files, including owner ,group changes, and also checks system integrity.[20]
Web services accessed by sending SOAP messages to endpoints. This is handed by transport layer security protocol such as HTTP, SSL, and TLS others. This ensures secured peer to peer messages. Web based security standards mapping to XML message security. All protocols use to carry security data as part of XML document. The XML document is critical part of security requirement of web services. [9]

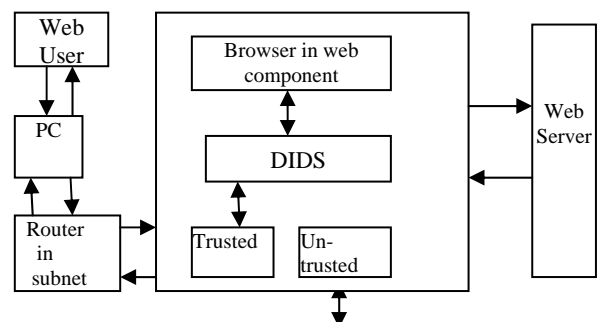

Figure.2. Dynamic Intrusion Detection System

**4.3 Secure Socket Layer and Transport Layer**
Security HTTP Secured Socket Layer Protocol: HTTP over Secured Socket Layer combination to secure communication between browser and web server systems. SSL Secured Socket Layer Protocol is transient, peer to peer communication, SSL protocol stack link associated with SSL session Record Protocol operation. These SSL sessions in association
between client & server by handshake protocol, with defined set of cryptographic parameters that may be shared by multiple SSL connections. [3, 12, 13]. HTTP protocol stack provides transfer information for web services interaction
can operate on top of SSL. Three layers are defined as part of SSL such as Hand shake protocol, The change of cipher spec protocol and the Alert protocol. These protocols are used in management of SSL exchange.

**4.4 Proposed Model: Dynamic Intruder Detection system**
The Automated Intruder Detection System shown in Figure. 2. It will detect the unauthorized or hacker requests by invoking a procedure Intrusion Detection System in four subpaths, that are user requests to subnet router point to point in Transport layer, browser in the Intrusion detection system detection system invokes a procedure to check, Certification, digital signature of trusted client. If trusted request as a result then it inserts the process for further processing into Deque. The deque holds a batch of trusted services routed to next hop via point to point protocol.In subpath3 browser contents security not known to attacker by pedlock security. The forth sub-path content in the web server connecting a session request for web services. The algorithm:1,2 and 3 depicted table. Intrusion Detection System Message Format alert:(messageid; create time;nt pstamp;date;time; source;node;address;message; flag)

4.4.1 Components of DIDS

Dynamic Intrusion Detection System has three components are sensor, Analyzer and User interface. Overall network security maintains a security state. when threat occurs by executing an event, the system will check the context of
the event and data by following

*4.4.1.1 Sensor*

Sensor are responsible for collecting data. Example network packet, log files, and system call traces, sensor collect and forward to the analyzer

*4.4.1.2 Analyzer*

Analyzer receiver input from one or more sensors from the system . Control the behavior of the system.

*4.4.1.3 User Interface*

The user interface to DIDS that enables a user to view output from the system or control the behavior of the system. System component as manager or console component.[20]

**Algorithm: 1** Initialization of Request
Procedure: DYNAMIC INTRUDER
DETECTON  SYSTEM
*Input :* Sensor/ node send a service request
*Output:* Trusted service or Un-trusted service
begin
1. Establish connection between sender and receiver
2. User system to web server consists of four sub paths
3. subpath1: user request to router in subnet(trusted system)
4. Subpath2:Web browser in router checks the procedure using Dynamic Intruder detection system.
5. Identify the request process trusted request pushed onto Deque and un-trusted requests rejected and access restricted.
6. Subpath3: Web browser content and security sign which are not known to the attacker. Ex icon with Padlock security sign unknown to the attacker.
7. Subpath4:Web content to Web server: Connecting via subnet routers with trusted systems the request connection established between sender to receiver
   using SSL handshake protocol
end

**Algorithm.2: Connection Establishment.**
Procedure PROCESS DETECTING TRUSTED REQUEST
*Input :*Web service request
*Output :* Secured HTTP Session layer
Begin
1. Read  DIDS
2. Client sends a request to Web server by invoking HandShake protocol using cryptographic parameters (clientid, clientMAC, ClientSecretKey, serverid)
3. During handshaking protocol session is created successfully by resuming the  new state if already the state is running. With Session identified by its state, prior to encryption algorithms.
4. Each connection creates a secure session layer and sets the flag. Here flag indicated the connection.
5. The request process is checking by Certificate Agent, and Digital certificate.
6. Detecting trusted service or un-trusted service. if request is trusted service then insert the process in Deque for further processing communication to next  hop if connection request is un-trusted requests are denied/ rejected
end

Algorithm.3 Closing the Connection
Procedure WS SECURED CONNCTION
*Input:* web service request message
*Output:* Secure access control by encryption
Begin
1. Read PROCESS DETECTING TRUSTED REQUEST
2. User data is verified with the data with existing data of concerned  web server.
3. If(HTTPrequest is successful) then Connection is established, under service access policy else Connection is closed with notification
4. if connection is established enable decryption of data at the web server
5. Message communication is accomplished by SSL encryption method.
6. close the connection
7. Connection closure if connection is closed in HTTP record
8. TLS level exchange close notify alert then close TCP connection
9. handle TCP close before alert exchange send or completed
End

## 5 Discussions and Interpretations

Secure Socket Layer provides security services between TCP and applications that use TCP. Internet standards TLS, SSL/TLS provide confidentiality using symmetric encryption
and message integrity and Authentication code. This DIDS (Dynamic Intruder Detection System) protecting the man-in the middle attacks and deny services. And allows trusted services forwarded to next hop to reach peer entity web server and Web service applications. The discussions and interpretations for web securities, precautions and remedies are described in table.2 provides the information related to web security/network security problems and proposed solutions/precautions to meet network securities QoS parameters such as confidentiality, integrity, data authentication, and availability of information to trusted users from web service systems to detect various attacks. The genuine merchants by Digital certificates and required policy constraints to validate authentication process in DIDS system architecture.

## 6 Conclusions

Web services are expected to play increasing important role for message communications over internet applications. Most of future work is web security. Online shopping and web services are increasing in the world. In this paper we described the fundamental concepts related to web security threats, web server architectures, web server protocols. QoS web service security is important concern in network communications. Firewalls security, various issues and challenges of web security. Discussed fundamental concepts, network security encryption and decryption process, and Network security hierarchies.

We proposed a novel Dynamic Intruder Detection System(AIDA) is safe guard against SSL secured transactions over message communications to intermediate routers that enable services to sender and receiver use Secured Session Layer protocol messages. This can be into three stages 1) Weak security assumption 2) Intruder attacks on browser 3)Trusted system detect service and safe guard information. As a case study we proposed the architecture of system in Figure .2.In future we can extend this paper to E-Commerce, Online Financial transactions, and this security concepts used for designing and developing Firewalls which will protect web services applications.